\documentclass[a4paper,11pt]{article}
\usepackage[english]{babel}
\usepackage[utf8]{inputenc}
\pdfoutput=1
\usepackage{braket}
\usepackage{physics}
\usepackage{color}
\usepackage{amssymb}
\usepackage{amsmath}
\usepackage{graphicx}
\usepackage{epstopdf}
\usepackage{subcaption}
\usepackage{comment}
\usepackage{mathtools}
\usepackage{ragged2e}
\usepackage{hyperref}
\hypersetup{
	colorlinks=true,
	linkcolor=blue,
	filecolor=cyan, 
	urlcolor=blue,
	citecolor=red,
} 
\usepackage[titletoc,toc,title]{appendix}
\numberwithin{equation}{section}
\usepackage{cleveref}
\usepackage[margin=2.5cm]{geometry}
\usepackage{cite}
\usepackage{epsfig}
\usepackage{float}
\usepackage{cancel}
\usepackage{ulem}
\usepackage{amsfonts}
\usepackage{enumitem}
\usepackage[font={footnotesize}]{caption}
\usepackage{authblk}
\usepackage{empheq}
\begin{document}

\title{Odd entanglement entropy in boundary conformal field theories and holographic moving mirrors}

\author{Anjali Kumari\thanks{\noindent E-mail:~ {\tt \href{mailto:anjali07kumariph@gmail.com} {anjali07kumariph@gmail.com}}}}

\author{Vinayak Raj\thanks{\noindent E-mail:~ {\tt \href{mailto:vraj@iitk.ac.in} {vraj@iitk.ac.in}}}}

\author{Gautam Sengupta\thanks{\noindent E-mail:~ {\tt \href{mailto:sengupta@iitk.ac.in} {sengupta@iitk.ac.in}}}}

\affil{
	Department of Physics,\\
	Indian Institute of Technology,\\
	Kanpur 208 016, India
}
\date{}
\maketitle

\thispagestyle{empty}

\begin{abstract}
In this article, we investigate the entanglement structure of bipartite mixed states in (1+1)-dimensional boundary conformal field theories (BCFT$_2$s) through the odd entanglement entropy (OEE) by employing an appropriate replica technique. In this regard we compare our results with the bulk entanglement wedge cross section (EWCS) for AdS$_3$ geometries with an end-of-the-world brane. We observe consistent extension of the holographic duality between the difference of the OEE and the entanglement entropy (EE) with the bulk EWCS, in the framework of AdS$_3$/BCFT$_2$ holography. Furthermore, we also extend our computations to the holographic moving mirror models where the Hawking radiation from eternal and evaporating black holes may be simulated depending on certain mirror profiles, and find consistent extension of the aforementioned holographic duality.

\end{abstract}

\clearpage

\tableofcontents

\clearpage

\section{Introduction} \label{sec:intro}
The characterization of quantum entanglement in quantum many body systems has become pivotal in investigating a wide range of phenomena, spanning from condensed matter physics to quantum gravity. In quantum information theory, the entanglement entropy (EE) has emerged as a useful measure to quantify the entanglement in bipartite states. The EE is defined as the von-Neumann entropy for the reduced density matrix corresponding to the bipartite quantum state. The computation of the EE is straightforward for finite quantum systems. However, for quantum many body systems, it becomes intractable due to infinite number of eigenvalues for the reduced density matrix in question. An appropriate replica technique to obtain the EE for such field theories was forwarded in \cite{Calabrese:2004eu, Calabrese:2009qy} where the authors constructed a replicated manifold to obtain the EE in certain replica limit.

The EE properly characterizes the entanglement for bipartite pure states only as it receives irrelevant classical and quantum contributions for mixed states. Thus a different correlation or entanglement measure is required to investigate the entanglement structure of bipartite mixed states. Several such computable quantities has been introduced in the literature, like the entanglement negativity \cite{Vidal:2002zz, Plenio:2005cwa}, the entanglement of purification \cite{Takayanagi:2017knl, Nguyen:2017yqw}, the reflected entropy \cite{Dutta:2019gen}, the balanced partial entanglement \cite{Wen:2021qgx}, etc. Another computable measure introduced in \cite{Tamaoka:2018ned} is termed as the \textit{odd entanglement entropy} (OEE) which may be interpreted as the von-Neumann entropy of a partially transposed density matrix. A replica technique was provided to obtain the OEE for (1+1)-dimensional field theories with conformal symmetry (CFT$_2$). In the context of AdS/CFT correspondence \cite{Maldacena:1997re, Witten:1998qj}, the authors in \cite{Tamaoka:2018ned} also proposed that the difference of the OEE and the EE in CFT$_d$s may be described holographically through the bulk entanglement wedge cross section (EWCS) in AdS$_{d+1}$ geometries. The OEE has further been explored in various contexts in \cite{Kusuki:2019evw, Mollabashi:2020ifv, Dong:2021clv, Ghasemi:2021jiy,  Basak:2022gcv, Basu:2023aqz}.

On a separate note, the study of CFTs on manifolds with boundaries, termed as boundary conformal field theories (BCFTs) \cite{Cardy:2004hm}, have gained importance in recent past due to its application in vast range of topics ranging from the physics of open strings and D-branes to boundary critical behaviour of condensed matter systems. In (1+1)-dimensions, it may be obtained by introducing a conformally invariant boundary on the complex plane, thus preserving one copy of the Virasoro symmetry algebra \cite{Cardy:2004hm, Sully:2020pza}. The holographic duality for such theories is dubbed as the AdS/BCFT correspondence \cite{Takayanagi:2011zk, Fujita:2011fp} where the bulk dual is described by an AdS space truncated by an end-of-the-world (EOW) brane with Neumann boundary condition. 

Recently in \cite{Akal:2020twv, Akal:2021foz}, the model of holographic moving mirrors in a BCFT$_2$ was employed to investigate the black hole information loss problem\footnote{Note that the moving mirror model has earlier been utilized to investigate the information loss problem in \cite{Hawking:1975vcx, Davies:1976hi, Birrell:1982ix, Good:2019tnf, Good:2020nmz}}. The authors considered certain profiles for the trajectory of a moving mirror in a complex plane which simulated the Hawking radiation from a black hole. The mirror acts as the boundary of the manifold and produces a radiation flux which may mimic the Hawking radiation from an eternal black hole, an evaporating black hole, or two evaporating black holes depending upon the mirror profile. In this setup, they reproduced the unitary Page curve for the entanglement entropy of a subsystem in the radiation flux of the moving mirrors. It is possible to map this moving mirror setup to a BCFT$_2$ on a half plane through a conformal transformation which simplifies the computations. Further interesting studies in this direction have been explored in \cite{Sato:2021ftf, Ageev:2021ipd, Reyes:2021npy, Kawabata:2021hac, BasakKumar:2022stg, Akal:2022qei, Good:2022wpw, Kumar:2023kse}

Noting the above developments, the study of the mixed state entanglement in BCFT$_2$s and its extension to the holographic moving mirror model becomes a natural point of enquiry. This was investigated in \cite{BasakKumar:2022stg} through the reflected entropy and the entanglement negativity. But the investigation of the OEE in this setup and the validity of its holographic duality remains an open question and is the main focus of this article. In this context, we obtain the OEE for bipartite adjacent and disjoint subsystems in BCFT$_2$s through an appropriate replica technique and compare the difference between the OEE and the EE with the corresponding EWCS obtained earlier in \cite{Lu:2022cgq, Basu:2023wmv, BasakKumar:2022stg}. We further extend our results to the holographic moving mirror setup by utilizing certain conformal transformations and again compare our results with earlier literature. Moreover, we plot the variation of the difference between the OEE and the EE with time to observe the transition between various phases, for mirror profiles simulating an evaporating and an eternal black hole.

The rest of the article is arranged as follows. In \cref{sec:review}, we provide a brief review of the several individual elements utilized in this study. Further in \cref{sec:OEE-BCFT}, we provide an appropriate replica technique and obtain the OEE for two adjacent and two disjoints subsystems in BCFT$_2$s in various phases. We also extend our computations to the moving mirror setup for all the cases discussed. Finally, in \cref{sec:summary}, we summarize our results and draw conclusions.

\section{Review of earlier literature}\label{sec:review}

\subsection{Odd entanglement entropy}\label{sec:OEE-review}
We begin with a brief review of the bipartite correlation measure termed as the odd entanglement entropy (OEE), introduced in \cite{Tamaoka:2018ned}. It could be understood as the von Neumann entropy of the partially transposed reduced density matrix of the given bipartite state. For a more precise definition, begin by considering a tripartite pure state composed of the subsystems $A_{1}$, $A_{2}$, and $B$. Subsequently, a bipartite mixed state is prepared by tracing out the degrees of freedom of the subsystem $B$ to obtain the reduced density matrix $\rho_{A_{1}A_{2}}$ defined on the Hilbert space $\mathcal{H} = \mathcal{H}_{A_{1}} \otimes \mathcal{H}_{A_{2}}$. The partial transposition of the reduced density matrix $\rho_{A_{1}A_{2}}$ with respect to the subsystem $A_{2}$ is then defined as
\begin{equation}\label{red}
	\mel{e^{(1)}_{i} {e^{(2)}_{j}} }{\rho^{T_{A_{2}}}_{A_{1}A_{2}}}{ e^{(1)}_{k} { e^{(2)}_{l}} } =  \mel{e^{(1)}_{i} { e^{(2)}_{l}}}{\rho_{A_{1}A_{2}}}{ e^{(1)}_{k} { e^{(2)}_{j}} } .
\end{equation}
where $\ket{e^{(1)}_{i}}$ and $\ket{ e^{(2)}_{j}}$ are the orthogonal bases for the Hilbert spaces $\mathcal{H}_{A_{1}}$ and $\mathcal{H}_{A_{2}}$ respectively. Further, the R\'enyi generalization of the OEE for the partially transposed density matrix is defined as follows
\begin{equation}\label{Renyi-OEE}
	S^{(n_{o})}_{o}(A_{1}:A_{2}) = \frac{1}{1 - n_{o}} \log\left[\Tr (\rho^{T_{A_{2}}}_{A_{1}A_{2}})^{n_{o}}\right],
\end{equation}
where ${n_{o}}$ is an odd integer\footnote{The trace in \cref{Renyi-OEE} may be obtained in a CFT$_2$ through a correlation function of twist fields $\Phi_n$ as will be discussed in \cref{sec:OEE-BCFT} in the context of boundary conformal field theories.}. The odd entanglement entropy $S_{o}$ for the given mixed state $\rho_{A_{1}A_{2}}$ may finally be obtained through the analytic continuation of the odd integer  ${n_{o}}$ $\rightarrow$ 1 in the above expression as follows
\begin{equation}\label{OEE-def}
	S_{o}(A_{1}:A_{2}) = \lim_{n_{o} \to 1}\left[S^{(n_{o})}_{o}(A_{1}:A_{2})\right].
\end{equation}

Moreover, in the context of the AdS/CFT correspondence, the difference between the OEE and the EE has been conjectured to be described holographically in terms of the bulk EWCS corresponding to the bipartite state under consideration, as follows \cite{Tamaoka:2018ned}
\begin{equation}\label{duality}
	S_{o}(A_{1}:A_{2}) - S(A_{1}{\cup}A_{2}) =  E_{W}(A_{1}:A_{2})
\end{equation}
where $S(A_{1}{\cup}A_{2})$ denotes the EE and $E_{W}(A_{1}:A_{2})$ denotes the EWCS for the subsystem $A_{1}{\cup}A_{2}$.

\subsection{Boundary conformal field theory} \label{sec:BCFT-review}
Boundary conformal field theory (BCFT) is a CFT defined on manifolds with a boundary. This may be defined in $(1+1)$-dimensions by considering a CFT$_2$ on the
half-plane $x \geq 0$. This BCFT$_2$ defined on the right half plane (RHP) preserves one copy of the Virasoro symmetry algebra of a usual CFT and it corresponds to a set of generators given by \cite{Sully:2020pza, Cardy:2004hm, Takayanagi:2011zk, Fujita:2011fp}
\begin{equation}
	\Tilde L_{n} = L_{n} + \bar L_{n},
\end{equation}
where $L_n$ and $\bar{L}_n$ are the generators of the two copies of Virasoro algebra of the usual CFT$_{2}$.

Any $n$-point correlation function in such BCFT$_2$s may be described by a $2n$-point correlator in the chiral CFT on the whole complex plane through the Cardy's \textit{doubling trick}\footnote{An alternative way to compute the BCFT correlator termed as the \textit{mirror method}, has been described in \cite{Akal:2021foz} where $\langle \sigma \sigma \rangle_\text{BCFT} \simeq \sqrt{\langle \sigma \sigma \sigma \sigma \rangle_\text{CFT}}$ with the four-point correlator being computed in the usual CFT (with both holomorphic and anti-holomorphic part) and not in the chiral CFT.} as follows \cite{Cardy:2004hm}
\begin{equation}
	\begin{aligned}
		{\langle \mathcal{O}_{h_{1} \bar h_{1}}(z_{1}, \bar z_{1}) \ldots \mathcal{O}_{h_{n} \bar h_{n}}(z_{n}, \bar z_{n}) \rangle}_\text{BCFT}
		\sim {\langle \mathcal{O}_{h_{1}} (z_{1}) \ldots \mathcal{O}_{h_{n}}(z_{n}) \mathcal{O}_{\bar h_{1}}(\bar z_{1})\ldots\mathcal{O}_{\bar h_{n}}(\bar z_{n}) \rangle}_\text{CFT} \,,
	\end{aligned}
\end{equation}
where $h_i,\bar {h}_i$ represent the conformal dimensions of the operator $\mathcal{O}$ at position $z_i$. This doubling trick leads to a non-vanishing one-point function for a scalar primary operator in the BCFT$_2$ which is given as
\begin{equation}
	{\langle \mathcal{O}_{h \bar h}(z, \bar z) \rangle}_\text{BCFT} = \frac{A_{\mathcal{O}}}{|z - z^{*}|^{\Delta}}  = \frac{A_{\mathcal{O}}}{|2y|^{\Delta}} ,
\end{equation}
where $z = x + i\,  y$, $\Delta = h + \bar h$ is the scaling dimension of the operator $\mathcal{O}$ and ${A_{\mathcal{O}}}$ is the normalization of the two-point function. Correspondingly, a two-point function on the half plane may be expressed in two different channels based on the location of the primary operators, here illustrated through twist-field operators, as follows
\begin{equation}
	\begin{aligned}
		\text{$\langle \Phi_n(z_1, \bar z_1) \bar {\Phi}_n(z_2, \bar z_2)\rangle_\text{BCFT} =\left\{
		\begin{array}{ll}
		\frac{g_b^{2(1-n)} \epsilon^{2\Delta_n}}{(4 y_1 y_2)^{\Delta_n}} ~~~~~~~~& \eta \to 0 ~~,\\\\
		 \frac{\epsilon^{2\Delta_n}}{|z_1 - z_2|^{2\Delta_n}} ~~~~~~~~& \eta \to 1 ~~,
		\end{array}\right.$}
	\end{aligned}
\end{equation}
where $\Delta_n$ is the scaling dimension of the twist field $\Phi_n$ and $g_b$ is the (regulated) partition function for the CFT on a disk with depend on the set boundary condition and is related to the boundary entropy \cite{Sully:2020pza}. The cross-ration $\eta$ in the previous expression is given by
\begin{equation}
	\eta = \frac{(z_1 - \bar z_1)(z_2 - \bar z_2)}{(z_1 - \bar z_2)(z_2 - \bar z_1)}\,.
\end{equation}

\subsection{Holographic moving mirror} \label{sec:MM}
In this subsection, we briefly review the holographic moving mirror model discussed in \cite{Akal:2020twv, Akal:2021foz}. The authors considered certain mirror trajectories described by the profile $x = Z (t)$ which mimic the Hawking radiation from eternal as well as evaporating black holes. The mirror acts as a boundary and the region to its right is described by a BCFT$_2$. This BCFT is mapped to a right-half-plane (RHP) through the following conformal transformation of the light cone coordinates $u = t - x$ and $v = t + x$,
\begin{equation}\label{static-map}
	\tilde{u} = p(u) ~~ , \qquad \tilde{v} = v ~~,
\end{equation}
where the function $p (u)$ is chosen such that the mirror becomes static $\tilde u - \tilde v = 0$, i.e., $v = p (u)$. In the original coordinates, this may be expressed as
\begin{equation}
	t + Z(t) = p(t - Z(t)) \,.
\end{equation}
The stress tensor in the tilde coordinates may be obtained by noting that the BCFT in the RHP is in the ground state. Thus, for the map in \cref{static-map}, the stress tensor is only given by the conformal anomaly as follows
\begin{equation}\label{stress-tensor}
	T_{uu} = \frac{c}{24} \left(\frac{3}{2} \left(\frac{p''(u)}{p'(u)}\right)^2 - \frac{p'''(u)}{p'(u)}\right) \, ,
\end{equation}
where primes represent derivatives with respect to $u$.

\begin{figure}[h!]
	\centering
	\begin{subfigure}[b]{0.45\textwidth}
		\centering
		\includegraphics[scale=0.4]{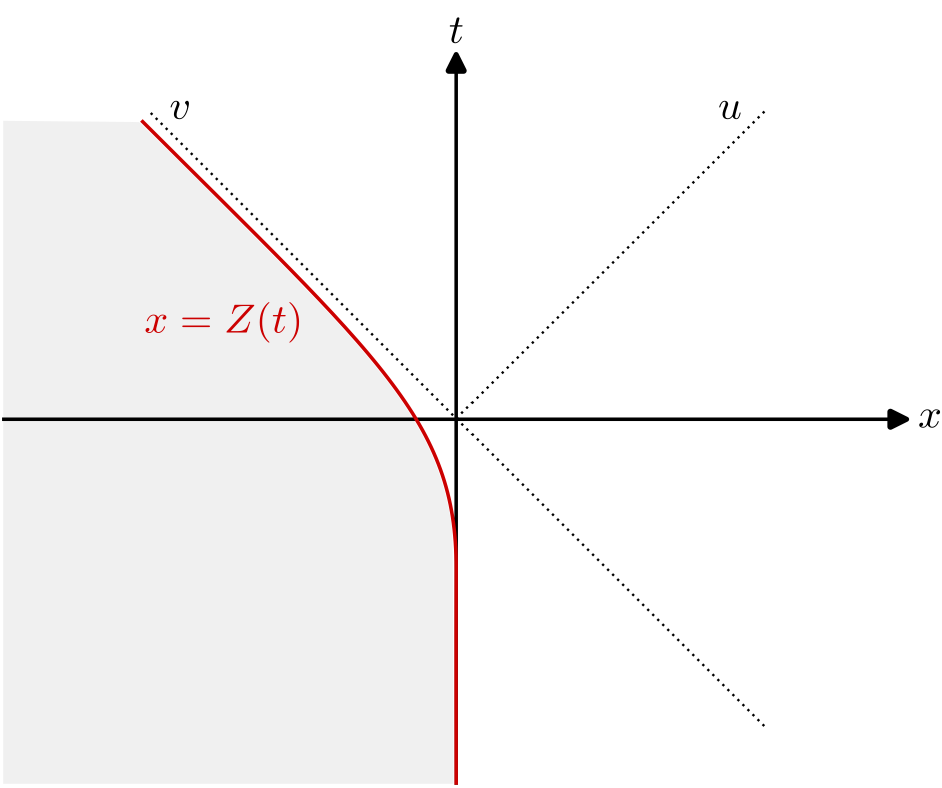}
		\caption{Schematics of the escaping mirror configuration modelling an eternal black hole.}
		\label{fig:escape}
	\end{subfigure}
	\hspace{0.8cm}
	\begin{subfigure}[b]{0.45\textwidth}
		\centering
		\includegraphics[scale=0.4]{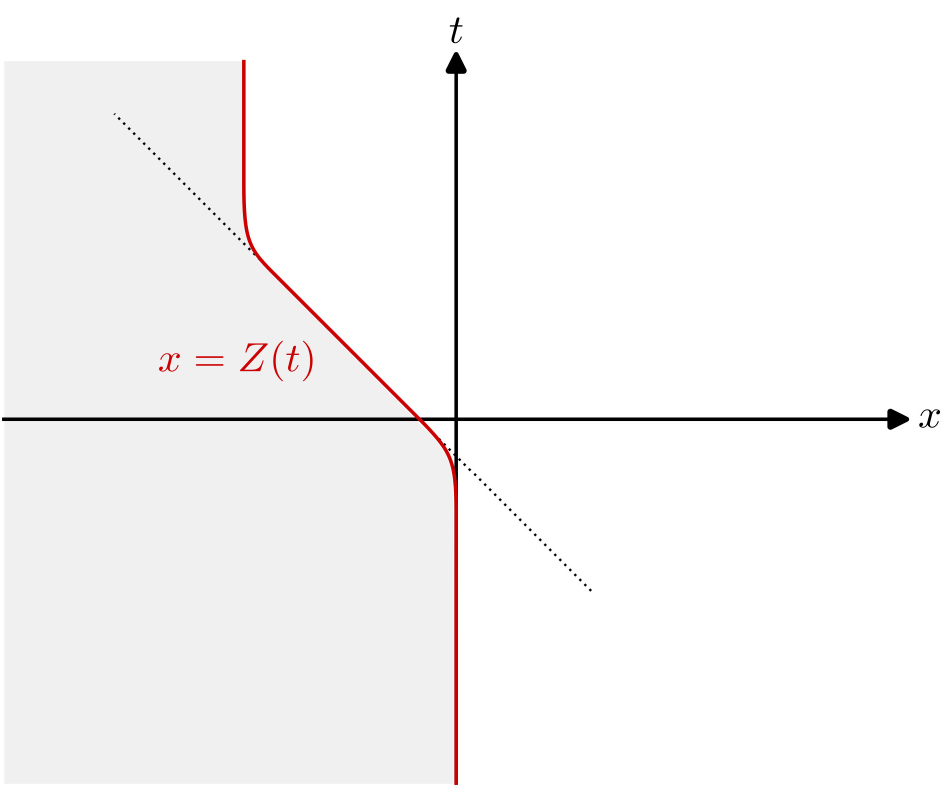}
		\caption{Schematics of the kink mirror configuration  modelling an evaporating black hole.}
		\label{fig:kink}
	\end{subfigure}
	\caption{Two mirror configurations simulating the Hawking radiation from black holes. The red curves describe the mirror trajectories.}
	\label{fig:mirror}
\end{figure}

As stated earlier, these holographic moving mirror setups could simulate the Hawking radiation from black holes. In particular, an eternal black hole with inverse temperature $\beta$ is modelled by considering an escaping mirror for which the profile is given by \cite{Akal:2020twv, Akal:2021foz}
\begin{equation}\label{esc-mirror}
	p (u) = - \beta \log \left( 1 + e^{-\frac{u}{\beta}} \right).
\end{equation}
As depicted in \cref{fig:escape}, in the early time limit $t \to 0$, the mirror is static $Z (t) \simeq 0$, whereas at late times $t \to \infty$, the mirror trajectory becomes almost null $Z (t) \simeq - t - \beta e^{-2 \frac{t}{\beta}}$.
On the other hand, an evaporating single-sided black holes is simulated by a kink mirror setup with the following profile \cite{Akal:2020twv, Akal:2021foz}
\begin{equation}\label{kink-mirror}
	p (u) = - \beta \log \left( 1 + e^{-\frac{u}{\beta}} \right) + \beta \log \left( 1 + e^{\frac{u - u_0}{\beta}} \right)\,,
\end{equation}
where $u_0 > 0$. In the large temperature limit $\beta \to 0$, the mirror trajectory is depicted in \cref{fig:kink} and may be approximated as
\begin{equation}
	\begin{aligned}
		\text{$Z (t) \simeq \left\{
			\begin{array}{ll}
				0 ~~~~~~~~~~& t < 0 ~~,\\
				-t ~~~~~~~~~~& 0 \leq t \leq \frac{u_0}{2} ~~,\\
				-\frac{u_0}{2} ~~~~~~~~~~& t > \frac{u_0}{2} ~~.
			\end{array}\right.$}
	\end{aligned}
\end{equation}

In the next section, we will now develop the replica technique for the OEE in BCFT$_2$ in order to investigate the nature of bipartite mixed state entanglement through the OEE in moving mirror models.

\section{OEE in BCFT$_{2}$}\label{sec:OEE-BCFT}

We will now establish the replica technique to obtain the OEE for bipartite states in BCFT$_2$s. To this end, consider the generic configuration of two disjoint subsystems $A_1 \equiv [x_{1},x_{2}]$ and $A_2 \equiv [x_{3},x_{4}]$ on a half-plane $x \geq 0$. Following \cite{Calabrese:2009qy, Calabrese:2012ew, Tamaoka:2018ned}, the trace $\Tr (\rho^{T_{A_{2}}}_{A_{1}A_{2}})^{n_{0}}$ of the partially transposed reduced density matrix in the R\'enyi OEE in \cref{Renyi-OEE}, may be obtained through a correlation function on twist field operators placed at the end points of the subsystems on the half-plane as follows
\begin{equation} \label{Tr-disj}
	\Tr (\rho^{T_{A_{2}}}_{A_{1}A_{2}})^{n_{0}} = {\langle {\mathcal{T}}_{n_{0}}(x_{1}) \bar {\mathcal{T}}_{n_{0}}(x_{2}) \bar {\mathcal{T}}_{n_{0}}(x_{3}) {\mathcal{T}}_{n_{0}}(x_{4}) \rangle}_{\text{BCFT}}\,.
\end{equation}
Here ${\mathcal{T}}_{n_{0}}$ and $\bar {\mathcal{T}}_{n_{0}}$ are the twist and anti-twist field operators in BCFT$_{2}$ respectively, with the scaling dimension $\Delta_{{\mathcal{T}}_{n_{0}}} $ as
\begin{equation} \label{conformal-weight-T}
	\Delta_{{\mathcal{T}}_{n_{0}}} = h_{{\mathcal{T}}_{n_{0}}} + \bar{h}_{\bar{\mathcal{T}}_{n_{0}}} = \frac{c}{12}\left(n_{0} - \frac{1}{n_{0}}\right) \,.
\end{equation}
\begin{figure}[h!]
	\centering
	\includegraphics[scale=01]{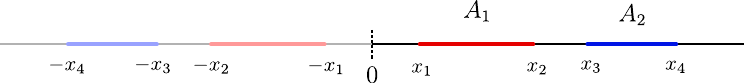}
	\caption{Two disjoint subsystems $A_{1}$ and $A_{2}$ in BCFT$_{2}$ with boundary at $x=0$. The darker region to the right ($x>0$) represent the BCFT$_2$ and the lighter region to the left ($x<0$) represent the copy obtained after utilizing the doubling trick.}
	\label{fig:disj}
\end{figure}

In the following subsections, we will now compute the OEE for two disjoint and two adjacent subsystems in BCFT$_2$s in different phases depending upon the location and size of the subsystems under consideration. We will further use the transformations in \cref{static-map} to map the results to the moving mirror setup.

\subsection{OEE for two disjoint subsystems} \label{sec:disj}

We will now compute the OEE for the mixed state configuration of two disjoint subsystems $A_1 \equiv [x_{1},x_{2}]$ and $A_2 \equiv [x_{3},x_{4}]$ in a BCFT$_2$ as shown in \cref{fig:disj}. In this scenario, the four-point function on the half-plane in \cref{Tr-disj} may be described by an eight-point correlator in the chiral CFT via the doubling trick, as follows
\begin{equation}\label{four-to-eight}
	\begin{aligned}
		&{\langle \mathcal{T}_{n_{0}}(x_{1}) \bar {\mathcal{T}}_{n_{0}}(x_{2})  \bar{\mathcal{T}}_{n_{0}}(x_{3}) {\mathcal{T}}_{n_{0}}(x_{4})\rangle}_{\text{BCFT}} \\
		&\qquad \qquad \quad \sim {\langle {\mathcal{T}}_{n_{0}}(x_{1}) \bar {\mathcal{T}}_{n_{0}}(x_{2}) \bar{\mathcal{T}}_{n_{0}}(x_{3}) {\mathcal{T}}_{n_{0}}(x_{4}) \bar {\mathcal{T}}_{n_{0}}(-x_{1}) {\mathcal{T}}_{n_{0}}(-x_{2})  {\mathcal{T}}_{n_{0}}(-x_{3}) \bar {\mathcal{T}}_{n_{0}}(-x_{4}) \rangle}_{\text{CFT}}\,.
	\end{aligned}
\end{equation}

In the large-$c$ limit, the above eight-point twist correlator factorizes based upon the sizes and the locations of the subsystems. We will now discuss several of these phases in the following.

\subsubsection{Phase-I} \label{sec:disj-I}
We begin by considering the two subsystems to be close to the boundary such that the separation between them is also small. In this scenario, the four-point BCFT correlator in \cref{four-to-eight} factorizes in the large-$c$ limit in the following way
\begin{equation}\label{four-bcft-factorize-1}
	\begin{aligned}
		{\langle {\mathcal{T}}_{n_{0}}(x_{1}) \bar {\mathcal{T}}_{n_{0}}(x_{2}) \bar{\mathcal{T}}_{n_{0}}(x_{3}) {\mathcal{T}}_{n_{0}}(x_{4})\rangle}_\text{BCFT} \sim {\langle {\mathcal{T}}_{n_{0}}(x_{1})\rangle}_\text{BCFT} {\langle \bar {\mathcal{T}}_{n_{0}}(x_{2})  \bar{\mathcal{T}}_{n_{0}}(x_{3}) \rangle}_\text{BCFT} {\langle {\mathcal{T}}_{n_{0}}(x_{4})\rangle}_\text{BCFT}\,.
	\end{aligned}
\end{equation}
Using the doubling trick, the one-point functions in the above expression may be computed easily. However, obtaining a general form for the two-point function which corresponds to a four-point twist correlator in the chiral CFT$_2$ is hard to compute. This computation is simplified in the large-$c$ limit for which the four-point correlator may be expanded in terms of the conformal blocks as follows
\begin{equation}\label{block-expansion}
	\begin{aligned}
		\frac{ {\langle {\mathcal{T}}_{n_{0}}(-x_{2}) \bar {\mathcal{T}}_{n_{0}}(x_{2}) \bar {\mathcal{T}}_{n_{0}}(x_{3}) {\mathcal{T}}_{n_{0}}(-x_{3}) \rangle}_\text{CFT}}{{(x_{2} - x_{3})}^{-\frac{c}{6}{\frac{n^{2} - 1}{n}}}} = \sum_p b_p \mathcal{F}(c,  h_{{\mathcal{T}}_{n_{0}}}, h_{p}, 1-x) \bar{\mathcal{F}}(c,  \bar h_{ \bar {\mathcal{T}}_{n_{0}}}, \bar h_{p}, 1-\bar x)
	\end{aligned}
\end{equation}
where $x = \frac{(2x_{2})(2x_{3})}{(x_{2} + x_{3})^{2}}$ is the cross-ratio, $b_p$ is the OPE coefficient, and $\mathcal{F}$ and $\bar{\mathcal{F}}$ are the Virasoro conformal blocks corresponding to the exchange of the primary operator with conformal dimension $h_{p}$. For the given configuration, the dominant contribution comes from the composite twist operator $\bar{\mathcal{T}}^{2}_{n_{0}}$ with the conformal weight $h_{{\mathcal{T}}_{n_{o}}}^{(2)} = h_{{\mathcal{T}}_{n_{o}}} = \Delta_{{\mathcal{T}}_{n_{0}}}/2$. The conformal block corresponding to this operator may then be obtained through the monodromy technique as follows
\begin{equation}\label{block}
	\log\mathcal{F}(c,  h_{{\mathcal{T}}_{n_{0}}}, h_{{\mathcal{T}}_{n_{o}}}^{(2)}, 1-x) = -h_{{\mathcal{T}}_{n_{o}}}^{(2)} \log \left[\frac{1 + \sqrt{x}}{1 - \sqrt{x}}\right]\,.
\end{equation}

The OEE may now be obtained for this phase by utilizing the above conformal block in \cref{four-bcft-factorize-1}, to be
\begin{equation}\label{OEE-disj-1}
	\begin{aligned}
		S_{o}(A_{1}:A_{2}) =&\frac{c}{6} \left[\log\bigg(\frac{2x_{1}}{\epsilon}\bigg)+ 2\log\left(\frac{x_{2} - x_{3}}{\epsilon}\right)+ \log\left(\frac{2x_{4}}{\epsilon}\right)\right]\\
		&\qquad \qquad \qquad \qquad \qquad \qquad +\frac{c}{6} \log \left[\left(\frac{\sqrt{x_{2}} + \sqrt{x_{3}}}{\sqrt{x_{2}} - \sqrt{x_{3}}}\right)\right] + 3S_\text{bdy}.
	\end{aligned} 
\end{equation}
where $S_\text{bdy}$ is the boundary entropy representing the entanglement with the boundary degrees of freedom. The corresponding entanglement entropy for this configuration may be obtained as
\begin{equation}\label{EE-disj-1}
	S(A_{1}{\cup}A_{2}) = \frac{c}{6} \left[\log\bigg(\frac{2x_{1}}{\epsilon}\bigg)+ 2\log\left(\frac{x_{2} - x_{3}}{\epsilon}\right)+ \log\left(\frac{2x_{4}}{\epsilon}\right)\right] + 2S_\text{bdy}. 
\end{equation}
The OEE modulo the contribution from the entanglement entropy may then be given by
\begin{equation}\label{EW-disj-1}
	\begin{aligned}
		S_{o}(A_{1}:A_{2}) - S(A_{1}{\cup}A_{2}) = \frac{c}{6} \log \left[\left(\frac{\sqrt{x_{2}} + \sqrt{x_{3}}}{\sqrt{x_{2}} - \sqrt{x_{3}}}\right)\right] + S_\text{bdy}.
	\end{aligned}
\end{equation}
which matches with the corresponding bulk EWCS computed earlier in \cite{Lu:2022cgq, Basu:2023wmv, BasakKumar:2022stg} and is consistent with the holographic duality in \cref{duality}. 

\subsubsection*{Moving mirror}
We will now compute the OEE for subsystems $A_1 \equiv [(t,x_1),(t,x_2)]$ and $A_2 \equiv [(t,x_3),(t,x_4)]$ in the moving mirror setup. To this end, we first generalize the result obtained in \cref{OEE-disj-1} to complex coordinates where the subsystems are located in the RHP at $\tilde{z}_i = (\tilde t_i, \tilde x_i)$ as follows
\begin{equation}\label{OEE-disj-1-complex}
	\begin{aligned}
		S_{o}(A_{1}:A_{2}) =&\frac{c}{6} \left[\log\bigg(\frac{\tilde z_{1} - \tilde z^{*}_{1}}{ \epsilon}\bigg)+\log\left(\frac{\tilde z_{2} - \tilde z_{3}}{ \epsilon}\right)+ \log\left(\frac{\tilde z^{*}_{2} - \tilde z^{*}_{3}}{ \epsilon}\right)+ \log\left(\frac{\tilde z_{4} - \tilde z^{*}_{4}}{ \epsilon}\right)\right]\\
		&\qquad \qquad \qquad \qquad \qquad \qquad \qquad \qquad \qquad \qquad \qquad + \frac{c}{12} \log \left[\frac{1 + \sqrt{\tilde \eta}}{1 - \sqrt{\tilde \eta}}\right] + 3S_\text{bdy}.
	\end{aligned} 
\end{equation}
where $\tilde z^*_i$ represents the complex conjugate of $\tilde z_i$ and the modified cross-ratio $\tilde \eta$ is given by 
\begin{equation}
	\tilde \eta =  \frac{(\tilde z^{*}_{2} - \tilde z_{2})(\tilde z_{3} - \tilde z^{*}_{3})}{(\tilde z^{*}_{2} - \tilde z_{3})(\tilde z_{2} - \tilde z^{*}_{3})}\,.
\end{equation}
Now utilizing the inverse of the static map in \cref{static-map}, the OEE in the moving mirror setup may be obtained as
\begin{equation}\label{OEE-disj-1-mm}
	\begin{aligned}
		S_{o}(A_{1}:A_{2}) = \frac{c}{6} \bigg[\log\bigg(&\frac{p(t - x_{1}) - t - x_{1}}{\epsilon} \bigg)+ \log\left(\frac{p(t - x_{2}) - p(t - x_{3})}{\epsilon}\right) + \log\left(\frac{  x_{2}  - x_{3}}{\epsilon}\right)\\
		& \qquad \qquad \quad + \log\left(\frac{p(t - x_{4}) - t - x_{4}}{\epsilon}\right)\bigg] + \frac{c}{12} \log \left[\frac{1 + \sqrt{\eta}}{1 - \sqrt{\eta}}\right]+ 3S_\text{bdy}\,,
	\end{aligned}
\end{equation}
where $p(u)$ represents the given mirror profile and the cross-ratio $\eta$ is given by
\begin{equation}
	\eta =  \frac{(t+x_2 - p(t - x_{2}))(p(t - x_{3}) - t-x_3)}{(t+x_2 -p(t - x_{3}))(p(t - x_{2}) - t-x_3)}\,.
\end{equation}
Subtracting the EE contribution from the OEE, we may obtain the following
\begin{equation}\label{EW-disj-1-mm}
	\begin{aligned}
		S_{o}(A_{1}:A_{2}) -S(A_1 \cup A_2)  =  \frac{c}{12} \log \left[\frac{1 + \sqrt{\eta}}{1 - \sqrt{\eta}}\right]+ S_\text{bdy}\,.
	\end{aligned}
\end{equation}
This matches with the corresponding moving mirror result for the EWCS in \cite{BasakKumar:2022stg}.

\subsubsection{Phase-II } \label{sec:disj-II}
In this phase, we consider that the two subsystems are still close to the boundary but the subsystem $A_1$ is small such that the eight-point chiral correlator in \cref{four-to-eight} factorizes into a four-point correlator and a pair of two-point correlators as follows,
\begin{equation}\label{eight-factorize-2}
	\begin{aligned}
		&{\langle {\mathcal{T}}_{n_{0}}(x_{1}) \bar {\mathcal{T}}_{n_{0}}(x_{2}) \bar{\mathcal{T}}_{n_{0}}(x_{3}) {\mathcal{T}}_{n_{0}}(x_{4}) \bar {\mathcal{T}}_{n_{0}}(-x_{1}) {\mathcal{T}}_{n_{0}}(-x_{2})  {\mathcal{T}}_{n_{0}}(-x_{3}) \bar {\mathcal{T}}_{n_{0}}(-x_{4}) \rangle}_{\text{CFT}}\\
		&\qquad \sim {\langle {\mathcal{T}}_{n_{0}}(x_{1}) \bar {\mathcal{T}}_{n_{0}}(x_{2}) \bar{\mathcal{T}}_{n_{0}}(x_{3}) \bar{\mathcal{T}}_{n_{0}}(-x_{1}) \rangle}_{\text{CFT}} {\langle {\mathcal{T}}_{n_{0}}(x_{4})\bar {\mathcal{T}}_{n_{0}}(-x_{4}) \rangle}_{\text{CFT}} {\langle {\mathcal{T}}_{n_{0}}(-x_{2}){\mathcal{T}}_{n_{0}}(-x_{3})\rangle}_{\text{CFT}} \,.
	\end{aligned}
\end{equation}
Note that similar factorization has also been observed in \cite{BasakKumar:2022stg} in the context of reflected entropy. For this case also, the four-point correlator may be obtained in the large-$x$ limit through the monodromy technique described earlier. Again, the dominant contribution arises from the twist operator $\bar{\mathcal{T}}_{n_0}^2$ for which the conformal block is as given in \cref{block}. However the cross-ratio $x$ for the four-point twist correlator in \cref{eight-factorize-2} in this case is modified to
\begin{equation}
	x = \frac{(x_{1} - x_{2})(x_{3} + x_{1})}{(x_{1} - x_{3})(x_{2} + x_{1})}\,.
\end{equation}
The OEE may now be obtained to be
\begin{equation}\label{OEE-disj-2}
	\begin{aligned}
		S_{o}(A_{1}:A_{2}) =& \frac{c}{6} \bigg[\log\bigg(\frac{2x_{1}}{\epsilon}\bigg)+ 2\log\left(\frac{x_{2} - x_{3}}{\epsilon}\right) + \log\left(\frac{2x_{4}}{\epsilon}\right)\bigg] \\
		& +\frac{c}{6}\log\left[\frac{x_{2}x_{3} - x_{1}^{2} + \sqrt{(x_{1}^{2}- x_{3}^{2})(x_{1}^{2} - x_{2}^{2})}}{x_{1}(x_{3}-x_{2})}\right] + 2S_\text{bdy} \, .
	\end{aligned} 
\end{equation}
The corresponding entanglement entropy for this case remains unchanged and is given by \cref{EE-disj-1}. The difference between the OEE and the EE may then be obtained as
\begin{equation}\label{EW-disj-2}
	S_{o}(A_{1}:A_{2}) - S(A_{1}{\cup}A_{2}) =\frac{c}{6}\log\left[\frac{x_{1}^{2} -x_{2}x_{3} + \sqrt{(x_{1}^{2}- x_{3}^{2})(x_{1}^{2} - x_{2}^{2})}}{x_{1}(x_{2}-x_{3})}\right] \, ,
\end{equation}
which matches with the corresponding EWCS obtained in \cite{Lu:2022cgq, Basu:2023wmv, BasakKumar:2022stg} and is consistent with the duality in \cref{duality} for the AdS$_3$/BCFT$_2$ scenarios. This provides a consistency check for our computation.

%{\color{blue}
%****************************************************************************\\
%These analytical results provide a comprehensive understanding of the system's properties in the large-c limit. By examining the factorization behavior and the influence of the logarithmic factors on the OEE, EE, and EWCS, we gain valuable insights into the intricacies of the mixed-state configuration in the $CFT_{1 + 1}$ system. The analysis underscores the importance of conformal blocks and the holographic duality in exploring the correlations and entanglement characteristics of the disjoint intervals within this fascinating physical system.
%
%****************************************************************************}

\subsubsection*{Moving mirror}
Generalizing the OEE obtained in \cref{OEE-disj-2} for complex coordinates $\tilde z = (\tilde t, \tilde x)$ in the RHP, we obtain
\begin{equation}\label{OEE-disj-2-complex}
	\begin{aligned}
	S_{o}(A_{1}:A_{2}) =&\frac{c}{6} \left[\log\bigg(\frac{\tilde z_{1} - \tilde z^{*}_{1}}{ \epsilon}\bigg)+\log\left(\frac{\tilde z_{2} - \tilde z_{3}}{ \epsilon}\right)+ \log\left(\frac{\tilde z^{*}_{2} - \tilde z^{*}_{3}}{ \epsilon}\right)+ \log\left(\frac{\tilde z_{4} - \tilde z^{*}_{4}}{ \epsilon}\right)\right]\\
	&\qquad \qquad \qquad \qquad \qquad \qquad \qquad \qquad \qquad \qquad \qquad + \frac{c}{6} \log \left[\frac{1 + \sqrt{\tilde \eta}}{1 - \sqrt{\tilde \eta}}\right] + 2S_\text{bdy}\,,
\end{aligned} 
\end{equation}
with the modified cross-ratio $\tilde \eta$ being given by
\begin{equation}
	\tilde \eta = \frac{(\tilde z_{1} - \tilde z_{2})(\tilde z_{3} - \tilde z_{1}^*)}{(\tilde z_{1} - \tilde z_{3})(\tilde z_{2} - \tilde z_{1}^*)}\,.
\end{equation}
We may now utilize the inverse of the static map in \cref{static-map} to obtain the OEE in the moving mirror setup for subsystems $A_1 \equiv [(t,x_1),(t,x_2)]$ and $A_2 \equiv [(t,x_3),(t,x_4)]$ as follows
\begin{equation}\label{OEE-disj-2-mm}
	\begin{aligned}
		S_{o}(A_{1}:A_{2}) = \frac{c}{6} \bigg[\log\bigg(&\frac{p(t - x_{1}) - t - x_{1}}{ \epsilon}\bigg)+ \log\left(\frac{p(t - x_{2}) - p(t - x_{3})}{ \epsilon}\right) + \log\left(\frac{  x_{2}  - x_{3}}{ \epsilon}\right)\\
		& \qquad \qquad \quad + \log\left(\frac{p(t - x_{4}) - t - x_{4}}{ \epsilon}\right)\bigg] + \frac{c}{6} \log \left[\frac{1 + \sqrt{\eta}}{1 - \sqrt{\eta}}\right]+ 2S_\text{bdy}\,,
	\end{aligned}
\end{equation}
where again $p(u)$ represents the given mirror profile for the moving mirror setup and the cross-ratio $\eta$ is given as
\begin{equation}
	\eta =  \frac{(p(t-x_1) - p(t - x_{2}))(p(t - x_{3}) - t-x_1)}{(p(t-x_1) - p(t - x_{3}))(p(t - x_{2}) - t-x_1)}\,.
\end{equation}
It is now possible to remove the EE contribution from the OEE to obtain the following expression
\begin{equation}\label{EW-disj-2-mm}
	\begin{aligned}
		S_{o}(A_{1}:A_{2}) - S(A_{1}{\cup}A_{2}) =  \frac{c}{6} \log \left[\frac{1 + \sqrt{\eta}}{1 - \sqrt{\eta}}\right]\,.
	\end{aligned}
\end{equation}
The above matches with the corresponding EWCS in moving mirror setup in \cite{BasakKumar:2022stg}.

\subsubsection{Phase-III} \label{sec:disj-III}
We now consider the case where the subsystems are close to the boundary but the subsystem $A_2$ is small. This leads to the following factorization of the eight-point chiral correlator in \cref{four-to-eight},
\begin{equation} \label{eight-factorize-3}
	\begin{aligned}
		&{\langle {\mathcal{T}}_{n_{0}}(x_{1}) \bar {\mathcal{T}}_{n_{0}}(x_{2}) \bar{\mathcal{T}}_{n_{0}}(x_{3}) {\mathcal{T}}_{n_{0}}(x_{4}) \bar {\mathcal{T}}_{n_{0}}(-x_{1}) {\mathcal{T}}_{n_{0}}(-x_{2})  {\mathcal{T}}_{n_{0}}(-x_{3}) \bar {\mathcal{T}}_{n_{0}}(-x_{4}) \rangle}_{\text{CFT}}\\
		&\qquad \sim {\langle \bar {\mathcal{T}}_{n_{0}}(x_{3}) {\mathcal{T}}_{n_{0}}(x_{4}) \bar{\mathcal{T}}_{n_{0}}(-x_{4}) \bar {\mathcal{T}}_{n_{0}}(x_{2})  \rangle_\text{CFT}}{\langle {\mathcal{T}}_{n_{0}}(x_{1})\bar {\mathcal{T}}_{n_{0}}(-x_{1})  \rangle_\text{CFT}} {\langle {\mathcal{T}}_{n_{0}}(-x_{2}){\mathcal{T}}_{n_{0}}(-x_{3})\rangle_\text{CFT}} \,.
	\end{aligned}
\end{equation} 
Similar to the previous subsection, the four-point twist correlator receives dominant contribution in the large-$c$ limit from the composite twist operator $\mathcal{T}_{n_0}^2$ for which the block is given in \cref{block}. The cross-ratio $x$ for the four-point twist correlator in \cref{eight-factorize-3} in this phase is given by
\begin{equation}
	x = \frac{(x_{3} - x_{4})(x_{2} + x_{4})}{(x_{3} + x_{4})(x_{2} - x_{4})} \,.
\end{equation}
We may now compute the OEE for this phase as follows
\begin{equation} \label{OEE-disj-3}
	\begin{aligned}
		S_{o}(A_{1}:A_{2}) =& \frac{c}{6} \bigg[\log\bigg(\frac{2x_{1}}{\epsilon}\bigg)+ 2\log\left(\frac{x_{2} - x_{3}}{\epsilon}\right) + \log\left(\frac{2x_{4}}{\epsilon}\right)\bigg]\\
		&  +\frac{c}{6}\log \left[\frac{x_{4}^{2} -x_{2}x_{3} + \sqrt{(x_{4}^{2}- x_{3}^{2})(x_{4}^{2} - x_{2}^{2})}}{x_{4}(x_{3}-x_{2})}\right] + 2S_\text{bdy}.
	\end{aligned} 
\end{equation}
Similar to the previous case, the entanglement entropy for this phase is same as given in \cref{EE-disj-1}. Consequently the difference between the OEE and the EE may be obtained as 
\begin{equation}\label{EW-disj-3}
	S_{o}(A_{1}:A_{2}) - S(A_{1}{\cup}A_{2}) = \frac{c}{6}\log \left[\frac{x_{4}^{2} -x_{2}x_{3} + \sqrt{(x_{4}^{2}- x_{3}^{2})(x_{4}^{2} - x_{2}^{2})}}{x_{4}(x_{3}-x_{2})}\right] \,.
\end{equation}
Note that the above result matches with the corresponding EWCS obtained in \cite{Lu:2022cgq, Basu:2023wmv, BasakKumar:2022stg} and is consistent with the holographic duality for the OEE in \cref{duality}.

\subsubsection*{Moving mirror}
In order to obtain the OEE for subsystems in moving mirror setup, we may now generalize the OEE obtained for this phase in \cref{OEE-disj-3} to complex coordinates for subsystems in the RHP at $\tilde{z}_i = ( \tilde{t}_i, \tilde{x}_i )$ to obtain an expression same as \cref{OEE-disj-2-complex} with the modified cross-ratio $\tilde{\eta}$ as
\begin{equation}
	\tilde \eta = \frac{(\tilde z_{3} - \tilde z_{4})(\tilde z_{2} - \tilde z_{4}^*)}{(\tilde z_{3} - \tilde z_{4}^*)(\tilde z_{4} - \tilde z_{2})}\,.
\end{equation}
Now, utilizing the inverse of the static map in \cref{static-map}, we may obtain the OEE in the moving mirror setup for bipartite state of two disjoint subsystems $A_1 \equiv [(t,x_1),(t,x_2)]$ and $A_2 \equiv [(t,x_3),(t,x_4)]$ as follows given in \cref{OEE-disj-2-mm} where again the cross-ratio $\eta$ is modified to
\begin{equation}
	\eta =  \frac{(p(t-x_3) - p(t - x_{4}))(p(t - x_{2}) - t-x_4)}{(p(t-x_3) - t - x_{4})(p(t - x_{4}) - p(t-x_2))}\,,
\end{equation}
where $p(u)$ represents the mirror profile. The OEE for this configuration modulo the entanglement entropy contribution may be given by
\begin{equation}
\begin{aligned}
	S_{o}(A_{1}:A_{2}) -S(A_1 \cup A_2) =\frac{c}{6} \log \left[\frac{1 + \sqrt{\eta}}{1 - \sqrt{\eta}}\right] \,.
\end{aligned} 
\end{equation}

\begin{figure}[ht]
	\centering
	\begin{subfigure}[b]{0.45\textwidth}
	\centering
	\includegraphics[scale=0.29]{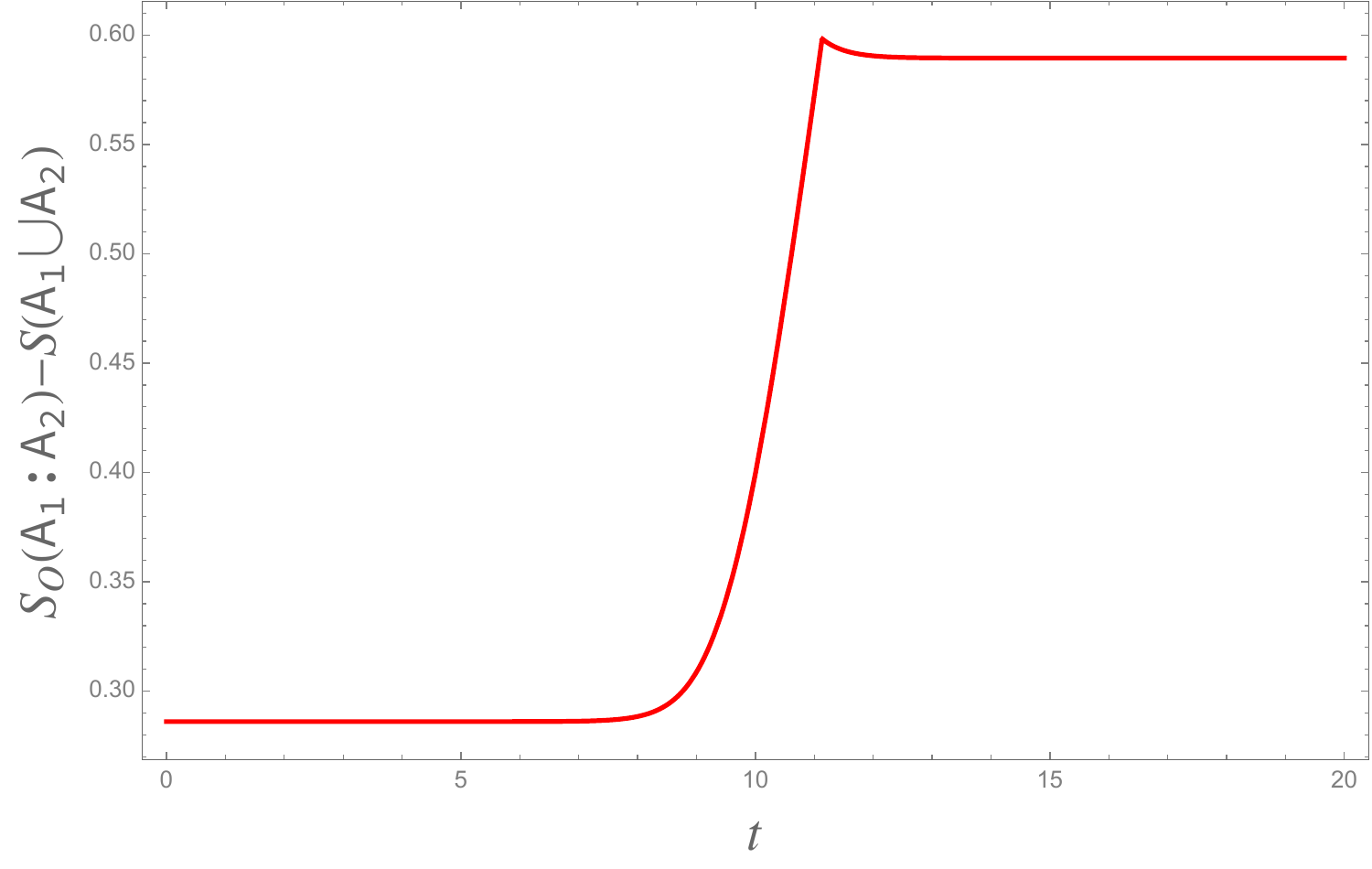}
	\caption{Escaping mirror: $A_1=[2,9]$ and $A_2=[10,20]$;\\ $c=1, \beta = 0.4, S_\text{bdy}=0.2$.}
	\label{fig:disj-escape}
	\end{subfigure}
	\hspace{0.8cm}
	\begin{subfigure}[b]{0.45\textwidth}
	\centering
	\includegraphics[scale=0.29]{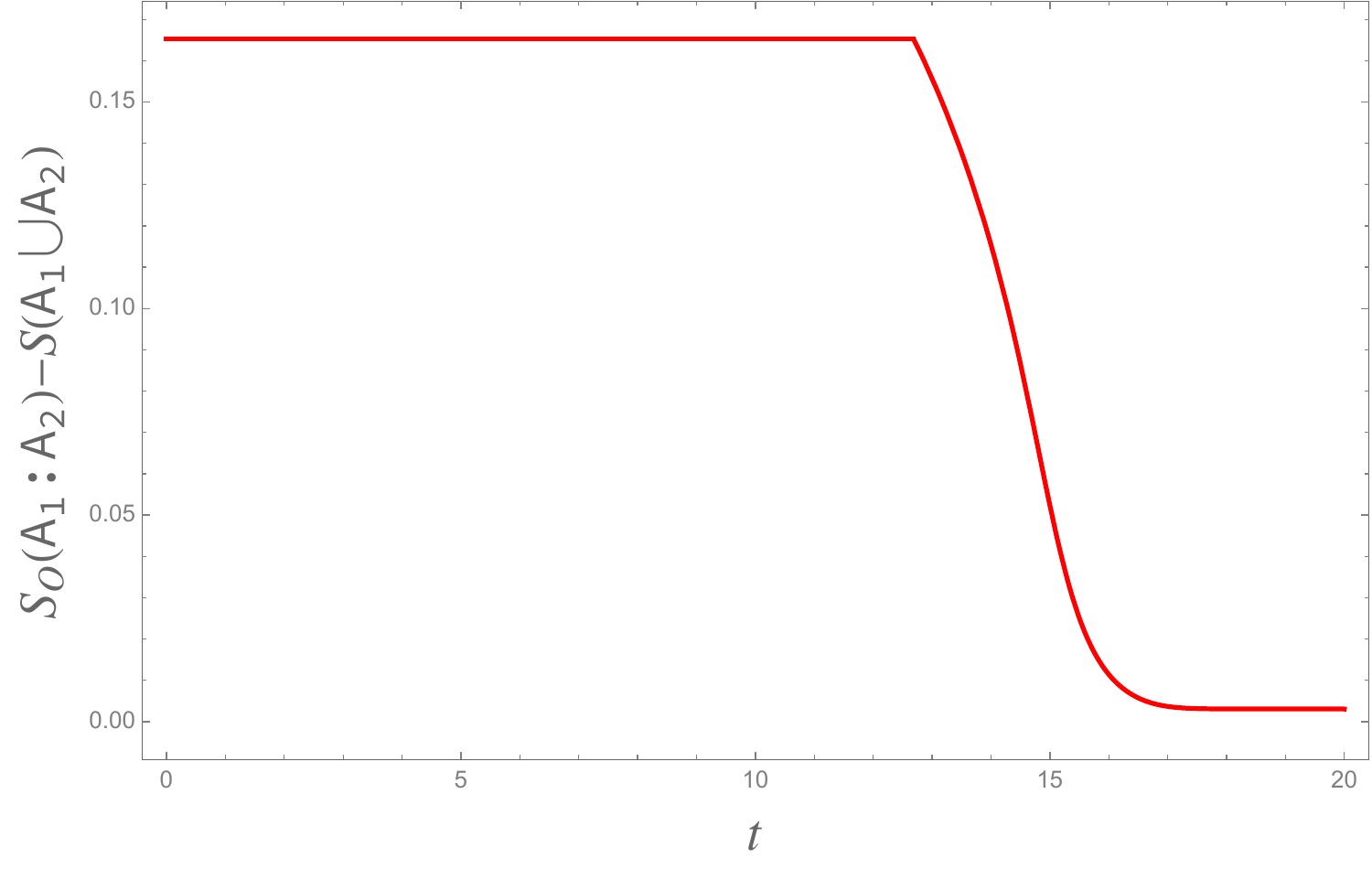}
	\caption{Kink mirror: $A_1=[12,15]$ and $A_2=[17,30]$;\\ $c=1, \beta = 0.25, S_\text{bdy}=0.2, u_0 = t - 20$.}
	\label{fig:disj-kink}
	\end{subfigure}
	\caption{Variation of the difference between the OEE and the EE  with time of two disjoint intervals for the escaping and the kink mirrors in BCFT$_2$.}\label{fig:disj-plot}
\end{figure}

We now plot the variation of the difference between the OEE and the EE in \cref{fig:disj-plot}. In \cref{fig:disj-escape}, we plot the difference between the OEE and the EE  for escaping mirror in a BCFT$_2$ described by the mirror profile in \cref{esc-mirror} which mimics an eternal black hole. As expected, the difference starts from a non-zero value at $t=0$ and starts rising. However after some time it transitions to a different phase, where it ultimately saturates to a constant value. Moreover, \cref{fig:disj-kink} depicts the kink mirror setup with the mirror profile given by \cref{kink-mirror} which simulates a evaporating black hole. The difference between the OEE and the EE start from a constant value but after some time transitions to a phase with declining difference, ultimately saturating to a vanishing value.

\subsection{OEE for two adjacent subsystems}  \label{sec:adj}
\begin{figure}[h!]
	\centering
	\includegraphics[scale=01]{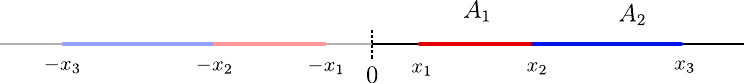}
	\caption{Two adjacent subsystems $A_{1}$ and $A_{2}$ in BCFT$_{2}$ described on half line $x\geq0$. The darker region on the right ($x>0$) represents the BCFT$_2$ and the lighter region on the left ($x<0$) represent the copy obtained after utilizing the doubling trick.}
	\label{fig:adj}
\end{figure}

In this subsection, we obtain the OEE for the mixed state of two adjacent subsystems $A_{1} \equiv [x_{1}, x_{2}]$ and $A_{2} \equiv [x_{2}, x_{3}]$ on the half-plane $x \geq 0$ as depicted in \cref{fig:adj}. This configuration may be obtained by considering the limit $x_2 \to x_3$ in \cref{Tr-disj} which results in the following three-point twist correlator
\begin{equation}\label{Tr-adj}
	\Tr (\rho^{T_{A_{2}}}_{A_{1}A_{2}})^{n_{0}} = {\langle {\mathcal{T}}_{n_{0}}(x_{1}) \bar {\mathcal{T}}^{2}_{n_{0}}(x_{2}) {\mathcal{T}}_{n_{0}}(x_{3}) \rangle}_{\text{BCFT}}\,.
\end{equation}
where the conformal dimensions $h_{\mathcal{T}_{n_0}} = \bar{h}_{\bar{\mathcal{T}}_{n_0}} = \bar{h}_{\mathcal{T}^2_{n_0}}$ of $\mathcal{T}_{n_{0}}$, $\bar{\mathcal{T}}_{n_{0}}$ and $\bar{\mathcal{T}}^2_{n_{0}}$ are as in \cref{conformal-weight-T}. Using Cardy's doubling trick, we may again express the above three-point twist correlator as a six-point function in the chiral CFT as follows
\begin{equation} \label{three-to-six}
  \begin{aligned}
      {\langle {\mathcal{T}}_{n_{0}}(x_{1}) \bar {\mathcal{T}}^{2}_{n_{0}}(x_{2}) {\mathcal{T}}_{n_{0}}(x_{3}) \rangle}_{\text{BCFT}} = {\langle {\mathcal{T}}_{n_{0}}(x_{1}) \bar {\mathcal{T}}^{2}_{n_{0}}(x_{2}) {\mathcal{T}}_{n_{0}}(x_{3}) \bar{\mathcal{T}}_{n_{0}}(-x_{1})  {\mathcal{T}}^{2}_{n_{0}}(-x_{2})  \bar{\mathcal{T}}_{n_{0}}(-x_{3}) \rangle}_{\text{CFT}}\,.
  \end{aligned}
\end{equation}
The explicit determination of the above six-point twist correlator in the chiral CFT is extremely challenging as it requires the knowledge of the full operator content of the theory. However, it may be evaluated in the large central charge limit for certain cases depending on the location of the subsystems $A_1$ and $A_2$ for which the correlation function in \cref{three-to-six} factorizes. In particular, we observe two different non-trivial phases which are discussed below.

\subsubsection{Phase-I} \label{sec:adj-I}
In this phase, we consider both the subsystems are close to the boundary $x=0$ and thus the six-point twist correlator in \cref{three-to-six} factorizes in three two-point correlators as follows
\begin{equation}
	\begin{aligned}
		&{\langle {\mathcal{T}}_{n_{0}}(x_{1}) \bar {\mathcal{T}}^{2}_{n_{0}}(x_{2}) {\mathcal{T}}_{n_{0}}(x_{3}) \bar{\mathcal{T}}_{n_{0}}(-x_{1})  {\mathcal{T}}^{2}_{n_{0}}(-x_{2})  \bar{\mathcal{T}}_{n_{0}}(-x_{3}) \rangle}_{\text{CFT}} \\
		&\qquad \qquad \qquad \quad \sim  {\langle {\mathcal{T}}_{n_{0}}(x_{1}) \bar {\mathcal{T}}_{n_{0}}(-x_{1})  \rangle}_{\text{CFT}}  {\langle  \bar {\mathcal{T}}^{2}_{n_{0}}(x_{2}) {\mathcal{T}}^{2}_{n_{0}}(-x_{2}) \rangle}_{\text{CFT}} {\langle {\mathcal{T}}_{n_{0}}(x_{3}) \bar {\mathcal{T}}_{n_{0}}(-x_{3})  \rangle}_{\text{CFT}}.
	\end{aligned}
\end{equation}
By using the standard form of two-point twist correlators in CFT$_2$s in the above expression, the OEE for this case may now be obtained through \cref{Renyi-OEE,OEE-def} as follows
\begin{equation}\label{OEE-adj-1}
	S_{o}(A_{1}:A_{2}) = \frac{c}{6} \bigg[\log\bigg(\frac{2x_{1}}{ \epsilon}\bigg)+ \log\left(\frac{2x_{2}}{ \epsilon}\right) + \log\left(\frac{2x_{3}}{ \epsilon}\right)\bigg] + 3 S_\text{bdy}\,.
\end{equation}
We can read off the entropy contribution for $A_1 \cup A_2$ in the above as follows
\begin{equation}\label{EE-adj-1}
	S(A_{1}{\cup}A_{2}) = \frac{c}{6} \left[\log\left(\frac{2x_{1}}{ \epsilon}\right)+ \log\left(\frac{2x_{3}}{ \epsilon}\right)\right] + 2S_\text{bdy}\,.
\end{equation}
Subtracting this entropy contribution from the OEE in \cref{OEE-adj-1}, we obtain
\begin{equation}\label{EW-adj-1}
	S_{o}(A_{1}:A_{2}) - S(A_{1}{\cup}A_{2}) = \frac{c}{6} \log\left(\frac{2x_{2}}{ \epsilon}\right) + S_\text{bdy}\,.
\end{equation}
which matches exactly with the correponding bulk EWCS computed in \cite{Lu:2022cgq, Basu:2023wmv, BasakKumar:2022stg}.

\subsubsection*{Moving mirror}
We will now generalize the above for complex coordinates where the subsystems are now located in the RHP at $\tilde z_i = (\tilde t_i, \tilde x_i)$. Following the above computation, the OEE for this configuration may be computed easily to be 
\begin{equation}\label{OEE-adj-1-complex}
	S_{o}(A_{1}:A_{2}) = \frac{c}{6} \bigg[\log\bigg(\frac{\tilde z_{1} - \tilde z^{*}_{1}}{ \epsilon}\bigg)+ \log\left(\frac{\tilde z_{2} - \tilde z^{*}_{2}}{ \epsilon}\right) +\log\left(\frac{\tilde z_{3} - \tilde z^{*}_{3}}{ \epsilon}\right)\bigg] + 3 S_\text{bdy}\,.
\end{equation}
where as earlier $\tilde z_i^*$ is the complex conjugate of $\tilde z_i$. Finally, inverting the static mirror map in \cref{static-map}, the OEE for subsystems $A_1 \equiv [(t, x_{1}),(t,x_{2})]$ and $A_2 \equiv [(t,x_{2}),(t,x_{3})]$ may be obtained to be
\begin{equation}\label{OEE-adj-1-mm}
	\begin{aligned}
		S_{o}(A_{1}:A_{2}) = \frac{c}{6} \Bigg[\log\bigg(\frac{p(t - x_{1}) - t - x_{1}}{ \epsilon \sqrt{p'(t - x_{1})}}\bigg)&+ \log\left(\frac{p(t - x_{2}) - t - x_{2}}{ \epsilon \sqrt{p'(t - x_{2})}}\right)\\
		&+ \log\left(\frac{p(t - x_{3}) - t - x_{3}}{ \epsilon \sqrt{p'(t - x_{3})}}\right)\Bigg] + 3 S_\text{bdy} \,,
	\end{aligned}
\end{equation}
where $p(u)$ represents the specific mirror profile and the square roots in the denominator arise from the transformation of the correlators to the moving mirror setup. The entropy contribution for $A_1 \cup A_2$ may be obtained in the present scenario to be
\begin{equation}\label{EE-adj-1-mm}
	\begin{aligned}
		S_{o}(A_{1}:A_{2}) = \frac{c}{6} \Bigg[\log\bigg(\frac{p(t - x_{1}) - t - x_{1}}{ \epsilon \sqrt{p'(t - x_{1})}}\bigg)+ \log\left(\frac{p(t - x_{3}) - t - x_{3}}{ \epsilon \sqrt{p'(t - x_{3})}}\right)\Bigg] + 2 S_\text{bdy} \,,
	\end{aligned}
\end{equation}
Subtracting this entropy contribution  from the OEE obtained in \cref{OEE-adj-1-mm}, we obtain
\begin{equation}\label{EW-adj-1-mm}
	\begin{aligned}
		S_{o}(A_{1}:A_{2}) - S(A_{1} \cup A_{2}) = \frac{c}{6}  \log\left(\frac{p(t - x_{2}) - t - x_{2}}{ \epsilon \sqrt{p'(t - x_{2})}}\right) + S_\text{bdy}\,.
	\end{aligned}
\end{equation}
which again matches with the corresponding bulk EWCS computed \cite{BasakKumar:2022stg}.

\subsubsection{Phase-II} \label{sec:adj-II}
In this phase, we consider that the subsystem $A_1$ is close to the boundary and is large in comparison to subsystem $A_2$ such that the three-point BCFT correlator in \cref{Tr-adj} factorizes as follows 
\begin{equation}\label{three-factor-BCFT}
	\begin{aligned}
		{\langle {\mathcal{T}}_{n_{0}}(x_{1}) \bar {\mathcal{T}}^{2}_{n_{0}}(x_{2}) {\mathcal{T}}_{n_{0}}(x_{3}) \rangle}_\text{BCFT} \sim {\langle {\mathcal{T}}_{n_{0}}(x_{1}) \rangle}_\text{BCFT} {\langle\bar {\mathcal{T}}^{2}_{n_{0}}(x_{2}) {\mathcal{T}}_{n_{0}}(x_{3}) \rangle}_\text{BCFT}\,.
	\end{aligned}
\end{equation}
The above two-point correlator on the half plane may be expressed in terms of a three-point twist field correlator in a CFT as follows \cite{BasakKumar:2022stg, Li:2021dmf}
\begin{equation}\label{two-to-three-cft}
	\begin{aligned}
		{\langle\bar {\mathcal{T}}^{2}_{n_{0}}(x_{2}) {\mathcal{T}}_{n_{0}}(x_{3}) \rangle}_\text{BCFT} \sim {\langle {\mathcal{T}}_{n_{0}}(-x_{3}) \bar {\mathcal{T}}^{2}_{n_{0}}(x_{2}) \bar {\mathcal{T}}_{n_{0}}(x_{3}) \rangle}_\text{CFT}\,.
	\end{aligned}
\end{equation}
The corresponding OEE for this configuration may now be computed by using \cref{Renyi-OEE,OEE-def} as follows
\begin{equation}\label{OEE-adj-2}
	{S_{o}(A_{1}:A_{2})} = \frac{c}{6} \bigg[\log\left(\frac{x_{3}-x_{2}}{\epsilon}\right) + \log\left(\frac{x_{2}+x_{3}}{\epsilon}\right) + \log\left(\frac{2 x_{3}}{\epsilon}\right) + \log\left(\frac{2 x_{1}}{\epsilon}\right)\bigg] + 2S_\text{bdy}\,.
\end{equation}
The entanglement entropy for this phase remains unchanged and is given by \cref{EE-adj-1}. The OEE excluding the contribution from the entanglement entropy may then be obtained as
\begin{equation}\label{EW-adj-2}
	\begin{aligned}
		{S_{o}(A_{1}:A_{2})} -S(A_1 \cup A_2) =\frac{c}{6} \log\left(\frac{x_{3}^2-x_{2}^2}{\epsilon \, (2x_3)}\right) \,,
	\end{aligned}
\end{equation}
which matches the corresponding EWCS \cite{Lu:2022cgq, Basu:2023wmv, BasakKumar:2022stg} apart from an additive constant. An explicit analysis of the OPE coefficient of the corresponding three-point twist correlator in \cref{two-to-three-cft} in lines of \cite{Dutta:2019gen} may lead to the recovery of this constant.

\subsubsection*{Moving mirror}
We now proceed to generalize the above computation for complex coordinates in the RHP with coordinates $\tilde{z} = (\tilde{t}, \tilde{x})$. The OEE for subsystems in this RHP may be obtained similarly as
\begin{equation}\label{OEE-adj-2-complex}
	\begin{aligned}
		S_{o}(A_{1}:A_{2}) = \frac{c}{6} \bigg[\log\left(\frac{z_{3} - z_{2}}{\epsilon}\right) + \log\left(\frac{z_{3} - z^{*}_{2}}{\epsilon}\right) + \log\left(\frac{z_{3} - z^{*}_{3}}{ \epsilon}\right)+\log\left(\frac{z_{1} - z^{*}_{1}}{ \epsilon}\right)\bigg] + 2 S_\text{bdy}.
	\end{aligned}
\end{equation}
where the endpoints of the subsystems are $\tilde{z}_i = (\tilde{t}_i, \tilde{x}_i)$  and $\tilde{z}_i^*$ are their complex conjugates. Again, inverting the static mirror map in \cref{static-map}, we may obtain the OEE for subsystems $A_1 \equiv [(t, x_{1}),(t,x_{2})]$ and $A_2 \equiv [(t,x_{2}),(t,x_{3})]$ in the moving mirror setup to be
\begin{equation}\label{OEE-adj-2-mm}
	\begin{aligned}
		S_{o}(A_{1}:A_{2}) = \frac{c}{6} \bigg[&\log\left(\frac{p(t - x_{3}) - p(t - x_{2})}{\epsilon \sqrt{p'(t - x_{2})p'(t - x_{3})}}\right) + \log\left(\frac{t + x_{3} - p(t - x_{2})}{\epsilon \sqrt{p'(t - x_{2})}}\right)\\ 
		& + \log\left(\frac{p(t - x_{3}) - t - x_{3}}{\epsilon \sqrt{p'(t - x_{1})}}\right) + \log\left(\frac{p(t - x_{1}) - t - x_{1}}{\epsilon \sqrt{p'(t - x_{1})}}\right)\bigg] + 2 S_\text{bdy}\,,
	\end{aligned}
\end{equation}
where, like earlier, the function $p(u)$ represents the specific mirror profile and the square root in the denominator is  due to the transformation of the corresponding correlation function to the moving mirror setup. The entanglement entropy for this case, as stated earlier, remains same as the previous case in \cref{EE-adj-1-mm}. Subtracting this entropy contribution, we may obtain the following
\begin{equation}\label{EW-adj-2-mm}
	\begin{aligned}
		{S_{o}(A_{1}:A_{2})} -S(A_1 \cup A_2) =\frac{c}{6} \bigg[\log\left(\frac{p(t - x_{2}) - p(t - x_{1})}{\epsilon \sqrt{p'(t - x_{2})p'(t - x_{1})}}\right) &+ \log\left(\frac{p(t - x_{2}) - t - x_{1}}{\epsilon \sqrt{p'(t - x_{2})}}\right)\\ 
		& - \log\left(\frac{p(t - x_{1}) - t - x_{1}}{\epsilon \sqrt{p'(t - x_{1})}}\right)\bigg] \,,
	\end{aligned}
\end{equation}
which matches the corresponding result for the EWCS in \cite{BasakKumar:2022stg} up to an additive constant.

\begin{figure}[H]
	\centering
	\begin{subfigure}[b]{0.45\textwidth}
		\centering
		\includegraphics[scale=0.29]{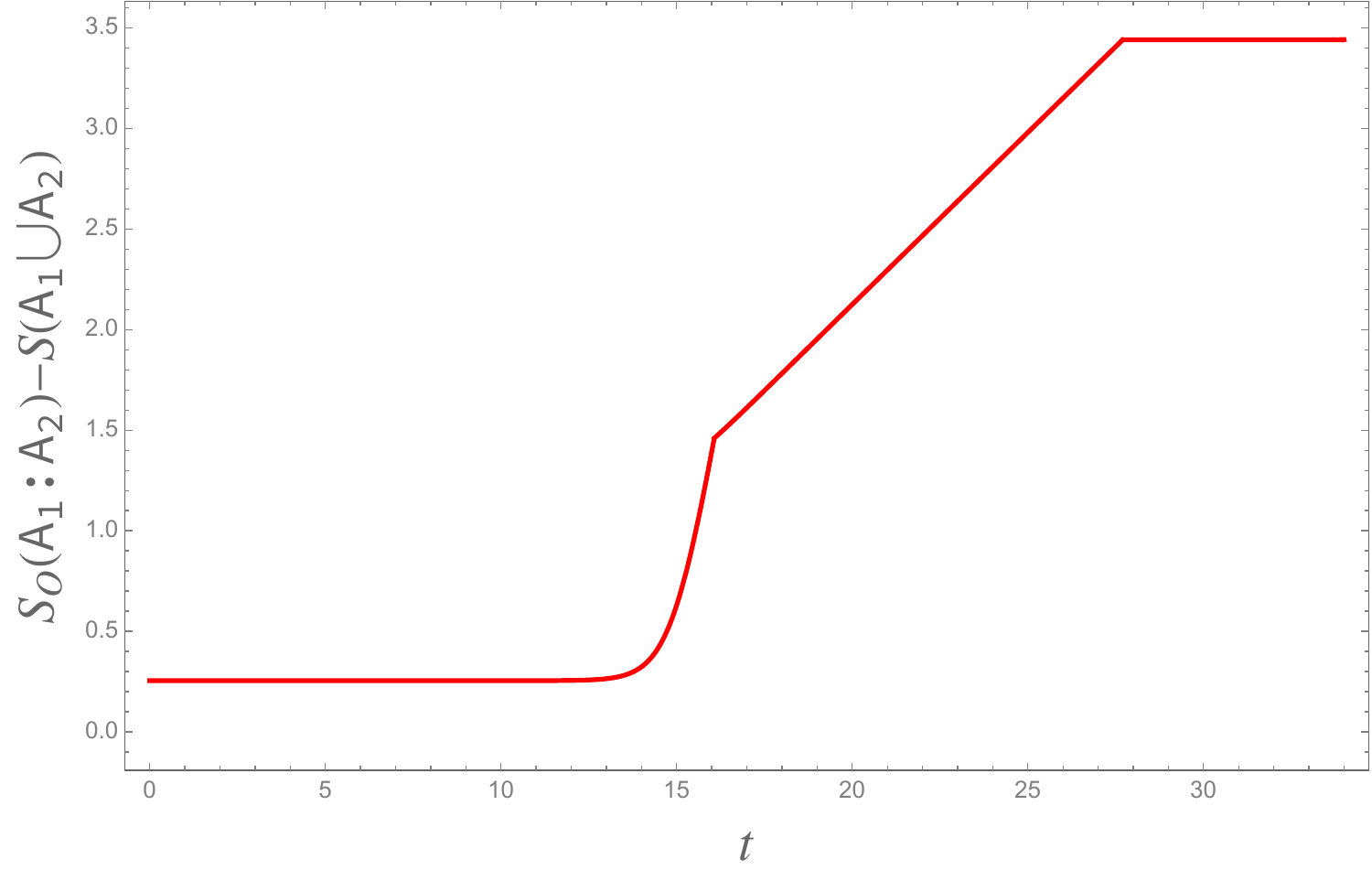}
		\caption{Escaping mirror: $A_1=[2,15]$ and $A_2=[15,18]$;\\ $c=1, \beta = 0.5, S_\text{bdy}=0.2, \epsilon=0.05$.}
		\label{fig:adj-escape}
	\end{subfigure}
	\hspace{0.8cm}
	\begin{subfigure}[b]{0.45\textwidth}
		\centering
		\includegraphics[scale=0.29]{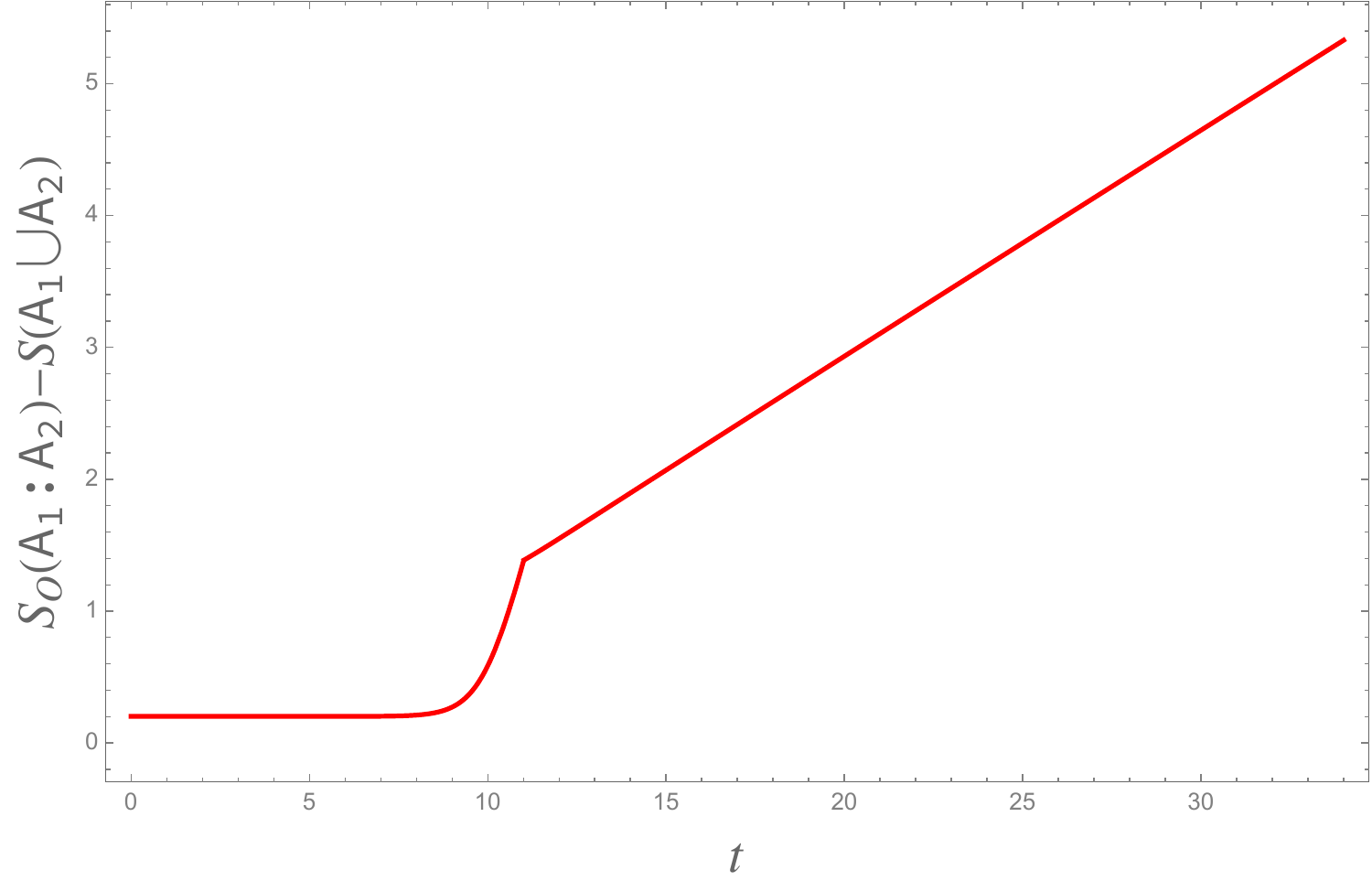}
		\caption{Kink mirror: $A_1=[2,10]$ and $A_2=[10,15]$;\\ $c=1, \beta = 0.5, S_\text{bdy}=0.2,\epsilon=0.05, u_0 = t - 20$.}
		\label{fig:adj-kink}
	\end{subfigure}
	\caption{Variation of the difference between the OEE and the EE  with time of two adjacent intervals for the escaping and the kink mirrors in BCFT$_2$.}\label{fig:adj-plot}
\end{figure}

We now plot the variation of the difference between the OEE and the EE in \cref{fig:adj-plot}. Fig. \ref{fig:adj-escape} depicts the difference between the OEE and the EE for two adjacent intervals for escaping mirror in a BCFT$_2$ with the mirror profile \cref{esc-mirror}. For the given values of the parameters in this case, the OEE transits from \hyperref[sec:adj-II]{phase-II} to \hyperref[sec:adj-I]{phase-I} at $t \approx 16$. However it returns back to \hyperref[sec:adj-II]{phase-II} at $t \approx 28$ and saturates at a constant value beyond that. Furthermore, \cref{fig:adj-kink} depicts the kink mirror setup with the mirror profile given by \cref{kink-mirror} which simulates an evaporating black hole. Again in this case, we observe a transition from \hyperref[sec:adj-II]{phase-II} to \hyperref[sec:adj-I]{phase-I} at $t \approx 11$. But unlike the escaping mirror case, the OEE stays in \hyperref[sec:adj-I]{phase-I} indefinitely beyond this time.

\section{ Summary and conclusions} \label{sec:summary}
In this article, we have investigated the mixed state entanglement structure of bipartite subsystems in BCFT$_2$s through the OEE by utilizing a suitable replica technique. We have compared our results with earlier literature to verify the validity of its holographic duality with the EWCS in the AdS$_3$/BCFT$_2$ framework. Furthermore we have extended our computations to the holographic moving mirror setup which simulates the Hawking radiation from a black hole. 

For the configuration of two disjoint intervals, we have utilized the doubling trick to describe the OEE for this configuration through an eight-point twist correlator in the chiral CFT. This eight-point correlator factorized in various ways in the large central charge limit, leading to three unique non-trivial phases for the OEE depending upon the location and size of the subsystems. We obtain the OEE for all these phases and found consistent matching of the difference between the OEE and the EE with the bulk EWCS, which provides a significant consistency check for our computations. We further generalize these computations to obtain the OEE for moving mirror setup. Moreover, we plot the variation of the difference between the OEE and the EE with time to observe the transition between the various phases. In particular we plot this variation for two specific mirror profiles mimicking an eternal black hole and an evaporating black hole.

We also consider the bipartite mixed state described by two adjacent subsystems in BCFT$_2$s. For this case, the OEE is characterized by a three-point BCFT twist field correlator, or a six-point chiral CFT correlator. Again, in the large-$c$ limit, this correlator may be factorized in various ways depending upon the subsystem size and location. In particular we observe two distinct non-trivial phases for which the OEE has been computed. The consistency of our computations have been verified by comparing the difference of the OEE and the EE with the bulk EWCS obtained earlier. We again extend the OEE to the case of moving mirrors. We finally plot the variation of the difference between the OEE and the EE for this configuration in moving mirror setup to observe transition between the two phases. 

In the present setup, the variations of the OEE with time is indicative of a rich phase structure of the mixed state entanglement in the Hawking radiation simulated by the moving mirrors which requires further investigations in other models of black hole evaporation like \cite{Balasubramanian:2020hfs, Balasubramanian:2020coy, Verheijden:2021yrb, Akers:2019nfi, Penington:2019kki, Balasubramanian:2021xcm, Afrasiar:2023nir, Basu:2022crn, Basu:2022reu}. A deeper understanding of this phase structure could be obtained by studying the methods of information recovery for mixed states from the black hole interior which seem to involve non-isometric encoding of the interior degrees of freedom \cite{Balasubramanian:2022fiy, Akers:2022qdl}. It would also be interesting to explicitly investigate the OEE for mixed states in island formalism by constructing an island formula for the OEE. We leave these exciting issues for future investigations.

\section*{Acknowledgement}
The authors would like to thank Debarshi Basu for several helpful discussions. The work of GS is partially supported by the Dr Jagmohan Garg Chair Professor position
at the Indian Institute of Technology, Kanpur.

\bibliographystyle{utphys}
\bibliography{reference}

\end{document}